\documentclass[11pt]{article}
\usepackage{moriond,epsfig}

\begin{document}
\title{
EXCITED STATES OF THE COOPER PROBLEM IN A THREE-DIMENSIONAL 
DISORDERED SYSTEM
}

\author{JOS\'E LAGES}

\address{Laboratoire de Physique Quantique, UMR 5626 du CNRS, 
Universit\'e Paul Sabatier, F-31062 Toulouse Cedex 4, France}

\maketitle
\abstracts{
This work presents the study of the excited states of the Cooper
problem in the three-dimensional Anderson model. It is shown
that the excited pair states remain localized while their excitation energy
$\Delta E$ is negative. For $\Delta E>0$ the particles are delocalized
over the three-dimensional lattice.
}

The Cooper problem \cite{cooper} is the cornerstone of the
well known BCS theory for superconductivity.
Even if the Cooper problem deals with only two interacting 
particles (TIP) above a frozen Fermi sea, it nevertheless
captures the essential physical features of the superconducting
states. Indeed, in comparison with the BCS theory, the solution 
of the Cooper problem leads to the appearance of coupled states 
with a qualitatively correct coupling energy and pair size.
Then it seems natural that the study of the Cooper problem
with disorder could provide a useful step in the understanding
of the superconductivity in presence of relatively strong
disorder. The first studies on the Cooper
problem with disorder were done for three-dimensional
\cite{lages1} and two dimensional \cite{lages2} systems.
These studies \cite{lages1,lages2} were focused on the properties of 
the ground state of two particles above a frozen
Fermi sea coupled via an on-site attractive Hubbard interaction.
This interaction creates a phase of bi-particle localized states (BLS)
in the regime of disorder where non-interacting states are delocalized.
At the same time the mean-field
solution of the Cooper problem (Cooper ansatz) gives delocalized
pairs. This shows that the non-diagonal matrix elements of interaction
play an important role in presence of disorder. These TIP results are
in qualitative agreement with recent many-body investigations 
\cite{bhargavi,lages3} of the ground state of the attractive Hubbard 
model with disorder.

Here the studies are concentrated on the properties of excited TIP
states.
For that purpose we introduce the Anderson Hamiltonian for one-particle,
\begin{equation}\label{H1}
H_1=\sum_{{\mathbf n}}E_{\mathbf n}
\left\arrowvert{\mathbf n}\right\rangle
\left\langle{\mathbf n}\right\arrowvert
+
V\sum_{\left\langle{{\mathbf{ n,m}}}\right\rangle}
\left\arrowvert{\mathbf n}\right\rangle
\left\langle{\mathbf m}\right\arrowvert
\end{equation}
where $\mathbf n$ and $\mathbf m$ are the index vectors on the
three-dimensional lattice with periodic conditions, 
$\left\langle ,\right\rangle$ denotes nearest neighbor sites, $V$
is the hopping term and the random on-site energies 
$E_{\mathbf n}$ are homogeneoulsy distributed in the energy interval
$\left[-\frac{W}{2},\frac{W}{2}\right]$, where $W$ is the disorder
strength. To study the effects of the attractive interaction ($U<0$)
on two particles
near the Fermi sea we generalize the Cooper approach\cite{cooper} 
for the disordered
case. We write the TIP Hamiltonian in the basis of one-particle
eigenstates of the Hamiltonian (\ref{H1}). In this basis
the Schr\"odinger equation for TIP reads
\begin{equation}\label{H2}
\left(E_{m_1}+E_{m_2}\right)\chi^{\lambda}_{m_1,m_2}
+U\sum_{m'_1,m'_2}Q_{m_1,m_2,m'_1,m'_2}\,\chi^{\lambda}_{m'_1,m'_2}
=E_{\lambda}\,\chi^{\lambda}_{m_1,m_2}.
\end{equation}
Here $E_{m}$ are the one-particle eigenenergies 
corresponding to the one-particle eigenstates 
$\left\arrowvert\phi_m\right\rangle$ of $H_1$
and $\chi^{\lambda}_{m_1,m_2}$ are the components 
in the 
non-interacting eigenbasis 
$\left\arrowvert\phi_{m_1}\otimes\phi_{m_2}\right\rangle$
of the 
$\lambda^{\mbox{\scriptsize th}}$ TIP eigenstate 
$\left\arrowvert\chi^\lambda\right\rangle$ corresponding
to the eigenenergy $E_{\lambda}$. The matrix elements 
$UQ_{m_1,m_2,m'_1,m'_2}$ give the interaction induced 
transitions between non-interactive eigenstates 
$\left\arrowvert\phi_{m_1}\otimes\phi_{m_2}\right\rangle$
and
$\left\arrowvert\phi_{m'_1}\otimes\phi_{m'_2}\right\rangle$.
These matrix elements are obtained by writing the Hubbard 
interaction in the non-interactive eigenbasis of the model
(\ref{H1}). In analogy with the original Cooper problem
\cite{cooper} the summation in (\ref{H2}) is done over 
the non-interacting states above the Fermi level,
in the labelling $E_{m'_{1,2}}>E_F$ corresponds to
$m'_{1,2}>0$. The Fermi energy is determined by 
a fixed filling factor $\nu=1/2$. To keep similarity with the Cooper
problem we restrict the summation on $m'_{1,2}$
by the condition $1<m'_1+m'_2<M$. In this way the
cut-off of M unperturbed orbitals introduces an effective
phonon frequency $\omega_D\propto M/L^3=1/\alpha$ 
where $L$ is the linear system size. When varying $L$
we keep $\alpha$ fixed so that the phonon frequency is 
independent of the system size. All the data in this work 
are obtained with $\alpha=30$ but we also checked that
the results are not sensitive to the change of $\alpha$.

In order to analyze the properties of the TIP excited 
states of the disordered Cooper 
problem we compute the energy level spacing distribution $P(s)$
obtained by the diagonalization of (\ref{H2}).
From $P(s)$ we calculate
\begin{equation}
\eta=\frac{\int_0^{s_0}\left[P(s)-P_{WD}(s)\right]ds}
{\int_0^{s_0}\left[P_P(s)-P_{WD}(s)\right]ds}.
\end{equation}
Here $P_P(s)=\exp(-s)$ and 
$P_{WD}(s)=\frac{\pi s}{2}\exp\left(-\frac{\pi s^2}{4}\right)$
are respectively the Poisson
and the Wigner-Dyson distributions, and $s_0\simeq 0.4729$
is their intersection point. In this way, $\eta$ varies
from $1$ [$P(s)=P_P(s)$] 
to 0 [$P(s)=P_{WD}(s)$]
 and 
thus characterizes a transition from localized to delocalized
states. For example for the one-particle problem this method
allows to detect efficiently the Anderson transition \cite{shklovskii}.

\begin{figure}[ht]
\begin{center}
\psfig{figure=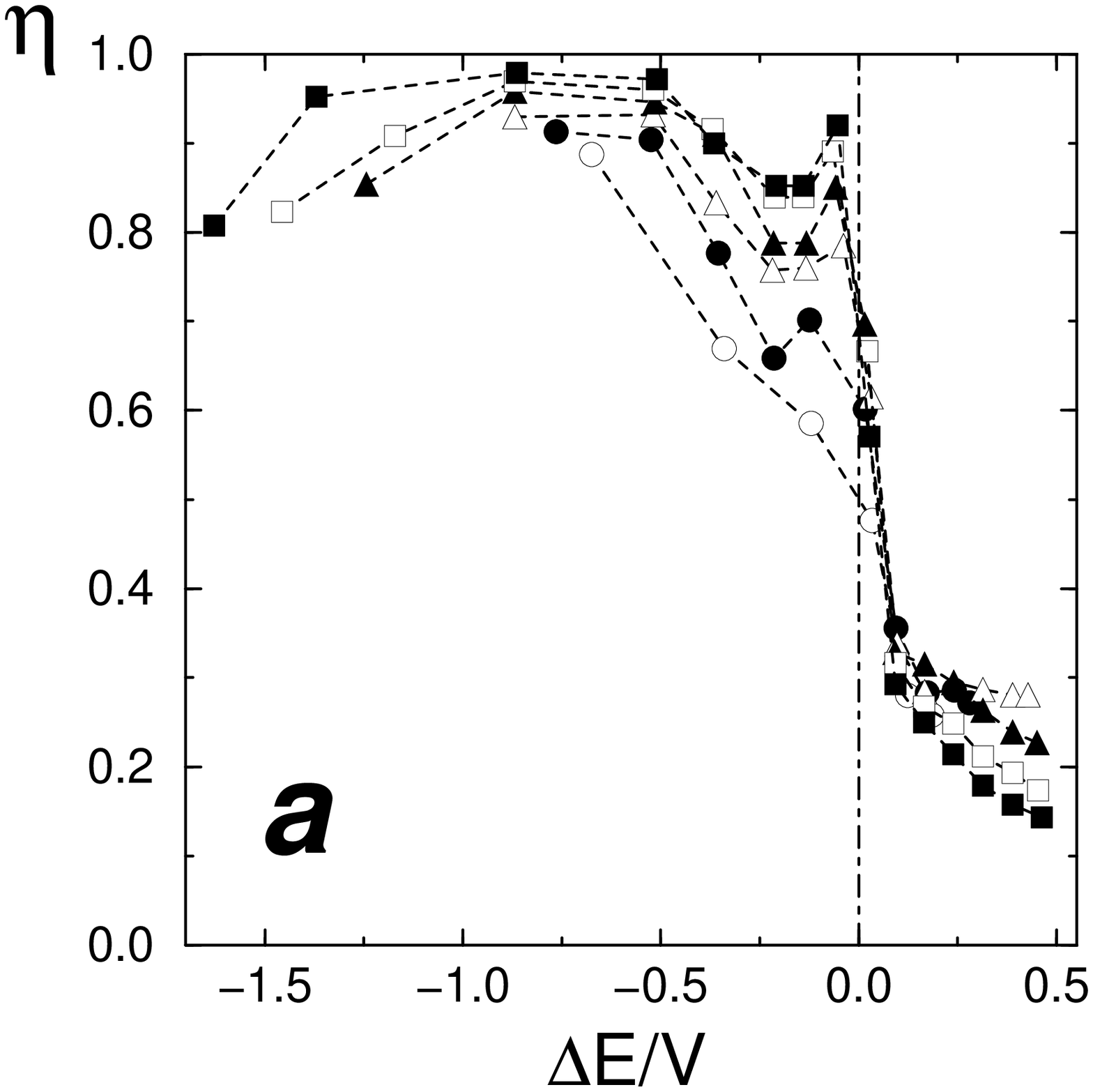,height=8cm,width=7.5cm}
\psfig{figure=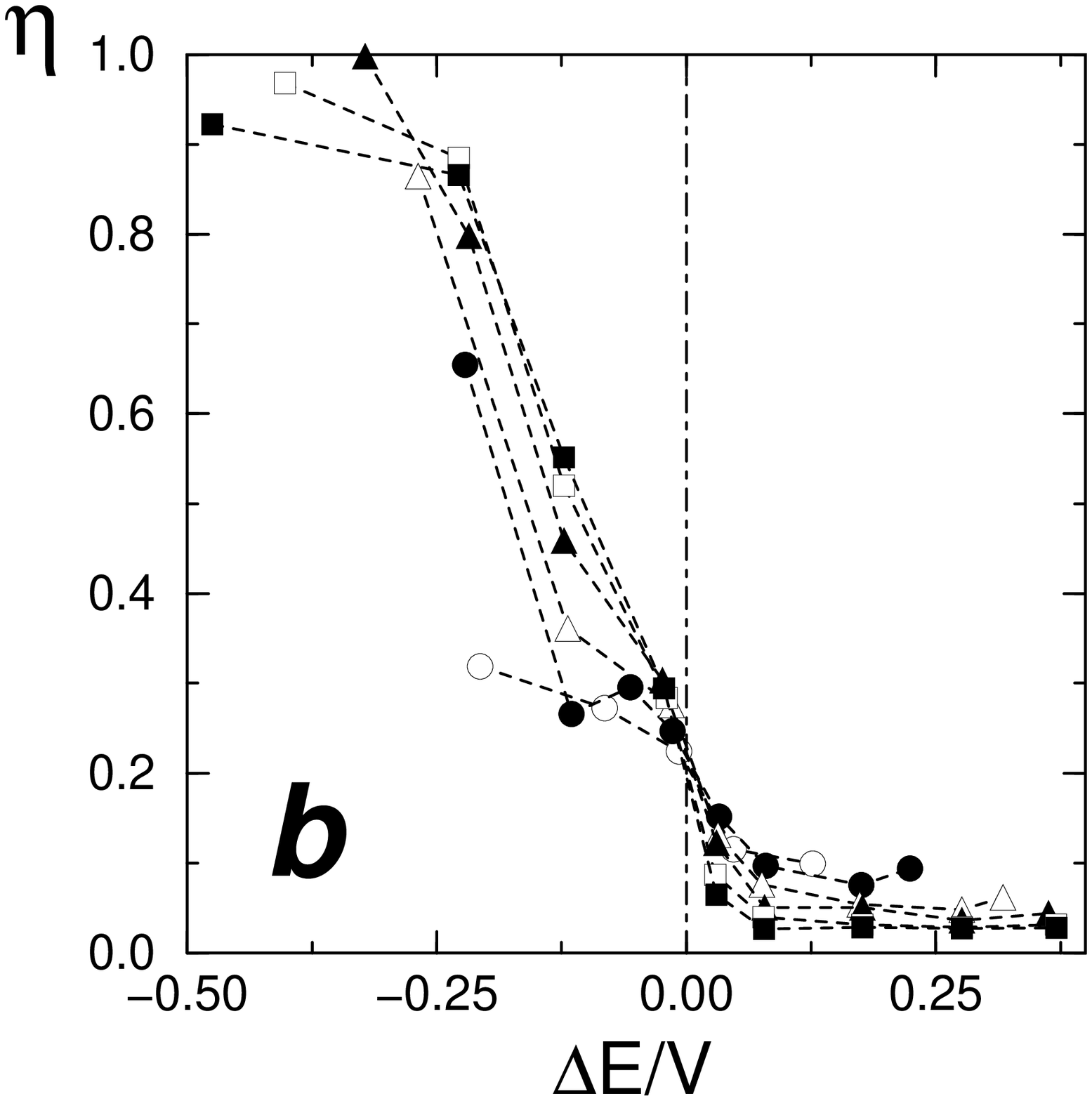,height=8cm,width=7.5cm}
\caption{
Dependence of $\eta$ on the excitation energy $\Delta E$ for 
({\bf a}) $W=18V>W_c\simeq 16.5V$
and ({\bf b}) $W=12V<W_c$. The interaction strength is $U=-4V$
and the linear lattice size is
$L=6\,(\circ),7\,(\bullet),8\,(\triangle),10\,(\mbox{full triangle})
,12\,(\mbox{square})$ and $14\,(\mbox{full square})$.
The number of disorder realization used is $N_D=3000$ and $\alpha=30$.
\label{fig1}
}
\end{center}
\end{figure}

\begin{figure}[ht]
\begin{center}
\psfig{figure=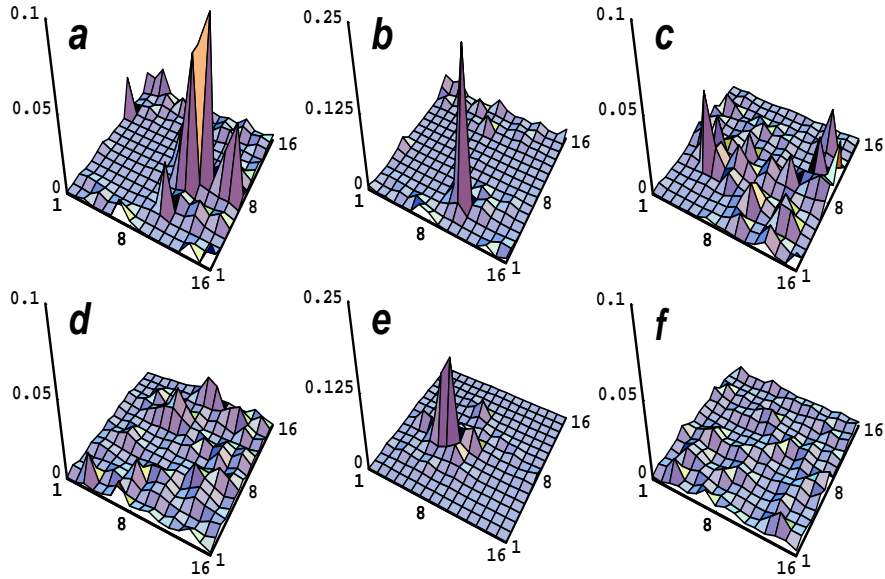,height=8cm,width=12cm}
\caption{
One-particle probability distribution
$f_\lambda({\mathbf n_1})=\sum_{{\mathbf n_2}}
\left\arrowvert\chi^\lambda_{\mathbf{n_1,n_2}}\right\arrowvert^2$
extracted from TIP eigenstates
$\left\arrowvert\chi^\lambda\right\rangle$ and projected on a (x,y)-plane.
The linear lattice size is $L=16$.
$f_\lambda({\mathbf n_1})$ is presented for $W=18V$ 
[
(a) ground state without interaction, 
(b) excited states $\lambda=25$ with $\Delta E_\lambda\simeq -0.25V$ 
for $U=-4V$
and 
(c) excited states $\lambda=758$ with $\Delta E_\lambda\simeq 0.25V$ 
for $U=-4V$
]
and for $W=12V$ 
[
(d) ground state without interaction, 
(e) excited states $\lambda=83$ with $\Delta E_\lambda\simeq -0.05V$ 
for $U=-4V$
and 
(f) excited states $\lambda=512$ with $\Delta E_\lambda\simeq 0.125V$ 
for $U=-4V$].
\label{fig2}
}
\end{center}
\end{figure}

In Fig. \ref{fig1}, $\eta$ is shown for disorder strength
$W=18V$ ({\bf a}) and $W=12V$ ({\bf b}), 
an interaction strength $U=-4V$ and for different linear 
lattice size, from $L=6$ up to $L=14$. $\eta$ is presented
versus the excitation energy $\Delta E=E-2E_F$ where $E$ is the 
energy of the
two interacting particles.
Without interaction
for $W=18V>W_c\simeq 16.5V$ \footnote{At the band center
$W_c\simeq 16.5V$ is the critical value of the disorder
at which the Anderson transition occurs\cite{shklovskii,japan}.}
the one particle states are well localized (see Fig. \ref{fig2}a).
When the on-site Hubbard attraction is switched on, coupled states
appear below the Fermi energy of two non-interacting particles
($\Delta E<0$). Fig. \ref{fig1}a clearly shows that at the 
thermodynamic limit the coupled states with $\Delta E<0$ are localized
($\eta\rightarrow 1$). As $W>W_c$ all the non-interacting orbitals
are localized in the lattice basis. The on-site attractive interaction acts 
as an additional constraint forcing the two particles (with opposite
spin) to stay together in the same well of potential (see Fig. \ref{fig2}b).
Above the Fermi
energy, $\eta$ tends slowly to 0 indicating  that these 
unbounded states ($\Delta E>0$) are delocalized. This delocalization
is due to the fact that an interaction (repulsive or attractive)
between particles destroies single particle localization and 
leads to a propagation of pairs of size $l_1$ over a distance
$l_c$ much larger than $l_1$ \,\,\cite{dima,imry}. The enhancement
factor $\kappa$ is then determined by the density $\rho_2$ of 
two-particle states 
coupled by the interaction and the interaction induced 
transition rate $\Gamma_2$ between noninteracting states, so that
$\kappa=\Gamma_2\rho_2$. For $\Delta E>0$ the density $\rho_2$
grows with the excitation energy ($\rho_2\propto \Delta E$)\cite{imry}
that strongly enhances the delocalization of pairs (see Fig. \ref{fig2}c).

In Fig. \ref{fig1}b, $\eta$ is shown for $W=12V<W_c$. For this regime
the non-interacting one-particle states are delocalized (see Fig. \ref{fig2}d).
For the different linear lattice size $L$ the ground state is 
still localized ($\eta\simeq 1$) in agreement with
\cite{lages1}. Indeed as 
$W/W_c\simeq 0.72>W_s/W_c\simeq 0.35$ the TIP ground state is 
in the BLS phase, where $W_s$ is the critical value of
the disorder for the superconductor-insulator transition
found in \cite{lages1} for TIP pairs. In the region
$-0.25V<\Delta E<0$ although $\eta$ drops from 0.9 to 0.25 a certain
localization of TIP pairs still remains (see Fig. \ref{fig2}e). For
$\Delta E>0$, the TIP states become delocalized (see Fig. \ref{fig2}f).

\begin{figure}[ht]
\begin{center}
\psfig{figure=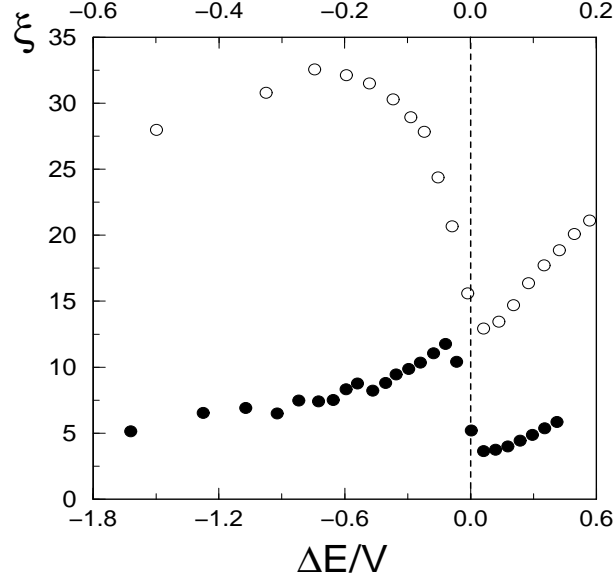,height=8cm,width=8cm}
\caption{
IPR $\xi$ as a function of the coupling energy $\Delta E$ for 
a disorder strength $W=12\,(\circ)$ [$\Delta E/V$ is then read
on the upper horizontal axis] and $W=18\,(\bullet)$ [$\Delta E/V$ is 
then read on the lower horizontal axis].
The interaction strength is $U=-4V$ and the linear lattice size $L=12$.
$N_D=300$ disorder realization are used, $\alpha=30$,
$2\epsilon\simeq 0.05$ for $W=18$ and $2\epsilon\simeq 0.025$ for 
$W=12V$. 
\label{fig3}
}
\end{center}
\end{figure}
In order to characterize the wave function properties of the 
excited states of the disordered Cooper problem, we compute
\begin{equation}
\xi\left(E\right)=\left\langle\left(
\sum_{m_1,m_2}\left\arrowvert\chi^\lambda_{m_1,m_2}\right\arrowvert^4
\right)^{-1}\right\rangle_{E_\lambda\in 
\left[E-\epsilon,E+\epsilon\right]}
\end{equation}
where the brackets mark the averaging over $N_D=400$ realizations of 
the disorder and over an energy interval $2\epsilon$ 
centered on $E$.
The inverse participation ratio (IPR) $\xi\left(E\right)$ counts the
average number of 
non-interacting states 
$\left\arrowvert\phi_{m_1}\otimes\phi_{m_2}\right\rangle$
occupied by the TIP wave functions belonging to energy interval
$\left[E-\epsilon,E+\epsilon\right]$. Fig. \ref{fig3} shows
the IPR as a function of the coupling energy $\Delta E$
for $W>W_c$ and $W<W_c$. For $W=18V$ as the one-particle noninteracting
states are localized, only few of these states ($\xi\simeq 5-10$)
are enough to build the TIP localized states with $\Delta E<0$.
On the contrary for $W=12V$ the TIP localized states ($\Delta E<0$)
occupy more noninteracting states ($\xi\simeq 30$)
than for $W=18V$ (Fig. \ref{fig3}).
Indeed, for $W<W_c$ 
the one-particle noninteracting states are delocalized and it is necessary
to have a sufficient number of these states to see the TIP eigenstates
localization induced by the quantum interferences.

In conclusion, the study of the excited states of the disordered Cooper
problem shows that the BLS phase found in 
\cite{lages1,lages2,bhargavi} exists for 
TIP eigenstates with energy $\Delta E<0$. This phase is destroyed for 
TIP eigenstates above the Fermi level $\Delta E>0$.

\section*{Acknowledgments}
I thank D.L Shepelyansky for useful and stimulating discussions.

\end{document}